\documentclass[journal=jctcce,manuscript=article]{achemso}

\usepackage[version=3]{mhchem} 
\usepackage[colorinlistoftodos]{todonotes} 
\usepackage{amsmath}
\usepackage{longtable} 
\usepackage{booktabs}
\DeclareMathOperator{\Tr}{Tr}

\author{Yanis R. Espinosa}
\affiliation[IFLySiB]
{Instituto de F\'isica de L\'iquidos y Sistemas Biol\'ogicos (CONICET-UNLP), Calle 59 Nro 789, 1900 La Plata, Argentina.}
\alsoaffiliation[Second University]
{Instituto Universitario de la Paz, Santa Lucía Km 14, Barrancabermeja, Colombia}

\author{H. Ariel Alvarez}
\affiliation[IFLySiB]
{Instituto de F\'isica de L\'iquidos y Sistemas Biol\'ogicos (CONICET-UNLP), Calle 59 Nro 789, 1900 La Plata, Argentina.}
\alsoaffiliation[Second University]
{Departamento de Ciencias Biol\'ogicas, Facultad de Ciencias Exactas, UNLP, Calle 47 y 115, 1900 La Plata, Argentina }

\author{Eduardo I. Howard}
\affiliation[IFLySiB]
{Instituto de F\'isica de L\'iquidos y Sistemas Biol\'ogicos (CONICET-UNLP), Calle 59 Nro 789, 1900 La Plata, Argentina.}
\alsoaffiliation[Second University]
{Department of Integrative Biology, Institut de G\'en\'etique et de Biologie Mol\'eculaire et Cellulaire, Centre de Biologie Int\'egrative, CNRS, INSERM, UdS, 1 rue Laurent Fries, 67404 Illkirch CEDEX, France.}

\author{C. Manuel Carlevaro}
\affiliation[IFLySiB]
{Instituto de F\'isica de L\'iquidos y Sistemas Biol\'ogicos (CONICET-UNLP), Calle 59 Nro 789, 1900 La Plata, Argentina.}
\alsoaffiliation[Second University]
{Universidad Tecnol\'ogica Nacional-FRBA, UDB F\'isica, Mozart 2300, C1407IVT Buenos Aires, Argentina.}
\altaffiliation{Corresponding author}
\email{manuel@iflysib.unlp.edu.ar.}
\phone{+54 221 425 4904 / 423 3283 }
\fax{+54 221 425 7317}

\title[Protein crystal]
  {Behavior of H-FABP-fatty acid complex in a protein crystal simulation}

\abbreviations{IR,NMR,UV}

\begin{document}

\begin{abstract}
Crystallographic data comes from a space-time average over all the unit cells within the crystal, so dynamic phenomena do not contribute significantly to the diffraction data. Many efforts have been made to reconstitute the movement of the macromolecules and explore the microstates that the confined proteins can adopt in the crystalline network. In this paper, we explored different strategies to simulate a heart fatty acid binding proteins (H-FABP) crystal starting from high resolution coordinates obtained at room temperature, describing in detail the procedure to study protein crystals (in particular H-FABP) by means of Molecular Dynamics simulations, and exploring the role of ethanol as a co-solute that can modify the stability of the protein and facilitate the interchange of fatty acids. Also, we introduced crystallographic restraints in our crystal models, according to experimental isotropic B{--}factors and analyzed the H-FABP crystal motions using Principal Component Analysis, isotropic and anisotropic B{--}factors. Our results suggest that restrained MD simulations based in experimental B{--}factors produce lower simulated B{--}factors than simulations without restraints, leading to more accurate predictions of the temperature factors. However, the systems without positional restraints represent a higher microscopic heterogeneity in the crystal. 
\end{abstract}

\section{Introduction}

X{--}ray crystallography has been the major contributor to our knowledge of the structure of macromolecules \cite{berman2013}. At the moment, almost 90\% of the structures in the Protein Data Bank \cite{berman2000} (PDB) have been solved by this technique, which has conditioned our way of representing proteins, offering its vision, as well as its limitations. The crystallographic data comes from a space-time average over all the crystal, so that the dynamic phenomena in an individual unit cell do not contribute significantly to the diffraction data, which are interpreted in terms of a mean structure \cite{kruschel2009}.

This single model representation is further reinforced by the fact that the crystal lattice prevents diffusion and restricts macromolecular movements \cite{andrec2007}. Many efforts have been made to \emph{reconstitute the movement} of the macromolecules and explore the microstates that the confined proteins can adopt in the crystalline network \cite{ma2015, janowski2013, kuzmanic2010}. Experimental approaches and different modeling techniques have been developed to recover this information \cite{wall2014, xue2014, li2014, kuzmanic2014x}. Among the computational tools, the use of Molecular Dynamics (MD) simulations and Normal Mode Analysis (NMA) \cite{terada2012, cerutti2010, kondrashov2007, meinhold2005} has introduced mayor advances. 

MD and NMA have the potential to recover information on dynamics and heterogeneity hidden in the X{--}ray diffraction data \cite{janowski2013}. Moreover, normal mode analysis offers an efficient way of modeling the conformational flexibility of protein structures \cite{kondrashov2007}. However, they could be hindered by the low quality of the structural model obtained by experimental data.

In this work, we analyzed a crystal of the heart fatty acid binding protein (H-FABP) based on the coordinates obtained by high-resolution X{--}ray and neutron diffraction techniques \cite{howard2016}. This protein is involved in the traffic of fatty acids inside the cell, and despite the extensive studies done in this family of proteins, the entry/release mechanism of the transported fatty acids are not well understood.

To study the behavior of the lipid/protein complex in the confined crystal form, we have explored different strategies in the setup of the Molecular Dynamics simulation. We describe here the procedure to build the initial crystal model, the influence of different solvents in crystal stability, and the tools to characterize the results in the context of exploring the dynamics of individual proteins in relation to conformational averaging.

\section{Computational methods}

\subsection{Molecular Dynamics}

When the structure of a macromolecule is solved by diffraction techniques, the positions of the atoms that have been identified in the asymmetric unit are deposited in the PDB, along with the information about crystallographic space group and its related symmetries. To model a crystal, it is necessary to use this symmetries to reconstruct the content of the unit cell and then, applying periodic boundary conditions, we are able to simulate an \emph{infinite}, borderless crystal.  

We obtained the crystal coordinates from an X{--}ray/neutron diffraction structure collected at room temperature (PDB ID: 5CE4). Using PyMOL \cite{delano2004} (\emph{symexp} command) we have applied the symmetry operations of the P2$_{1}$2$_{1}$2$_{1}$ space group to the protein and all structural water molecules identified (crystallographic waters). The values for the unit cell dimensions were $a =$ 3.4588 nm, $b =$ 5.5307 nm, $c =$ 7.1283 nm.
Considering that the length of the \emph{X} axis is close to the cut-off used during the simulation, we doubled the cell in this direction to avoid self-influence across periodic boundary conditions, and by this way the initial box dimensions were $2\times1\times1$ of $2L_{X} =$ 6.9176 nm, $L_{Y} =$ 5.5307 nm, $L_{Z} =$ 7.1283 nm. Hence, the simulation box contained eight H-FABP molecules, each complexed with a fatty acid (four complexes per unit cell) and 3769 SPC/E water molecules \cite{berendsen1987} from which 1376 were crystallographic water molecules. Four of the eight proteins contain palmitic acid, 
and the other four contain oleic acid (\emph{i.e.}, 4 H-FABP--palmitic acid complexes and 4 H-FABP--oleic acid complexes in the simulation box).

The effective pH was assumed to be 7.5, same as in the crystallization buffer. The protonation status of individual Asp, Glu, Lys, Arg, and His residues was obtained by PROPKA \cite{olsson2011} calculations for H-FABP in a crystal-lattice environment, leading to a charge of $-$1 per H-FABP molecule.
Thus, the net charge of each H-FABP{--}fatty acid complex was $-$2, so sixteen Na$^{+}$ counterions were added to neutralize the total charge of the system. 
The system was simulated using the united-atom GROMOS 54A7 force field \cite{Schmid2011}. Parameters for topologies of palmitic and oleic acid were obtained from Tsfadia and cols. (2007)\cite{tsfadia2007} and from POPC (1-palmitoyl-2-oleoyl-sn-glycero-3-phosphocholine) parametrization for this force field, and added to it (see Figure S1 and topologies incorporated as Supporting Information for details on their parametrization). 

The energy of the simulated system was initially minimized following a process where we applied 500 steps of steepest descent algorithm until a potential energy gradient $\Delta$E $\leq$ 1000 kJ mol$^{-1}$ was achieved. The protein atoms being harmonically restrained to their initial positions with a force constant of 25,000 kJ mol$^{-1}$ nm$^{-2}$ in all Cartesian directions. After assigning random initial velocities from a Maxwell-Boltzmann distribution at 100 K, the system was subsequently heated in three steps of 50 K and one step of 43 K, up to 293 K, simulating during 100 ps for each step. Simultaneously, for the same time lapse, the atomic position restraints in each protein molecule were uniformly relaxed down to zero (harmonic potential force constant relaxed from 25,000 to 0 kJ mol$^{-1}$ nm$^{-2}$ in steps of 5,000 kJ mol$^{-1}$ nm$^{-2}$).
The C$\alpha$ atoms from residues with a temperature factor (B{--}factor) lower than a value near 10 (44 atoms) were kept restrained throughout these equilibration runs using a restraining elastic constant of 25,000 kJ mol$^{-1}$ nm$^{-2}$ (see Table 1S in Supporting Information for details of B{--}factor values for each atom). The equilibration runs were performed at constant volume.

After equilibration, three different schemes at 293 K were applied for the treatment of the crystal unit cell volume and the deformations on the lattice: 

\begin{itemize}
\item NVT with C$\alpha$ atoms restraint,
\item NVT without restraints and,
\item NpT without restraints.
\end{itemize}
 
 The production simulations were run for 500 ns for each scheme using the GROMACS 2016.3 \cite{abraham2015} biomolecular simulation package with a 2 fs integration step. During equilibration and production, protein and non-protein groups were coupled separately to a heat bath using the Velocity{--}rescale thermostat \cite{bussi2007} with a relaxation time of 0.05 ps. In the NpT ensemble simulations, the pressure was calculated using a Parrinello{--}Rahman barostat \cite{parrinello1981} at 1 bar with a relaxation time of 1.0 ps. The bond lengths were constrained using LINCS algorithm \cite{hess1997} while electrostatic interactions were computed using the Particle Mesh Ewald method \cite{abraham2011opt}. A cut-off of 1.2 nm was applied both for the van der Waals and Coulomb interactions with a Verlet cut-off scheme. All calculations were carried out on a Linux server Intel Core i7-6700 3.40 GHz eight Core Processor with a NVIDIA GeForce GTX 1080 GPU. 

\subsection{Role of ethanol on Protein-Ligand interaction}

With the aim of assessing the role of ethanol in the dynamics of fatty acid exchange in confined proteins (\emph{i.e.}, in a protein crystal), we performed MD simulations of the same system in an aqueous solution of ethanol with an ethanol:water ratio of 1:37 (the same ratio used in the fatty acid exchange experiments of this system). Identical protocol of minimization and stabilization as in the Protein-Ligand crystal in water was used for this new system. After stabilization, the crystal was simulated for 500 ns, keeping always restrained only the C$\alpha$ from residue Ile114 in each protein with the initial force constant (25,000 kJ mol$^{-1}$ nm$^{-2}$). This residue was chosen for a number of reasons, namely, its low isotropic B{--}factor, an almost spherical anisotropic B{--}factor and its long distance from the relevant sites in terms of global movements of the protein. Simulation was performed at constant volume and also at 293 K.

\subsection{Essential dynamics}

\paragraph{Principal component analysis.} Collective coordinates, as obtained by a principal component analysis (PCA) of atomic fluctuations, are commonly used to predict a low-dimensional subspace in which essential protein motion is expected to take place \cite{daidone2012}. An atomic covariance matrix based on fluctuations of main-chain atoms is diagonalized to generate eigenvectors and eigenvalues that describe collective modes of fluctuation of the positions of the atoms in the protein \cite{Amadei1993}. Sorting the eigenvectors by the size of the eigenvalues shows that the configurational space can be divided in a low dimensional \emph{essential} subspace in which most of the positional fluctuations are confined. \cite{daidone2005} Thus, by PCA method, each element of the covariance matrix $C$ can be represented as \cite{Amadei1993, maisuradze2010}:

\begin{equation}
 \label{eq:eq_1}
 C_{i,j} = \langle x_{i} - \langle x_{i}\rangle\rangle \langle x_{j} - \langle x_{j} \rangle\rangle ,
\end{equation}

where $x_{1}, \ldots, x_{3N}$ are the mass-weighted Cartesian coordinates of an $N-$particle system and $\langle \rangle$ represents the average over all instantaneous structures sampled during the simulations.
The symmetric $3N\times 3N$ matrix $C$ can be diagonalized with an orthonormal transformation matrix $T$, $q = T^{T} (x - \langle x \rangle)$, which transforms $C$ into a diagonal matrix $\Lambda = \langle q~q^{T} \rangle$ of eigenvalues $\lambda_{i}$:
\begin{equation}
 \label{eq:eq_2}
  \Lambda = T^{T}CT = \text{diag} (\lambda_{1}, \lambda_{2}, \ldots , \lambda_{3N})
\end{equation}

	where $\lambda_{1} \geq \lambda_{2} \geq \ldots \geq \lambda_{3N}$. 
The $i$th column of $T$ is the eigenvector belonging to $\lambda_{i}$. Thus, the MD trajectory can be projected on the eigenvectors to determine the principal components (PC) $q_{i}(t), i = 1, \ldots, 3N $. 

The first few PCs typically describe collective global motions of the system, with the first PC	containing the largest mean-square fluctuation. Our covariance matrix was calculated using the C$\alpha$ carbons from the H-FABP crystal during the total time of the trajectory for each scheme simulated. 

\subsection{B--Factors Calculation}

In order to further analyze the behavior of the crystal simulation, we performed the theoretical calculation of the isotropic and anisotropic B{--}factors (\emph{i.e.},the mean-square displacements of the atoms, also termed anisotropic displacement parameters - ADPs) for the simulation runs so as to compare them with their experimental values. They can be obtained from the Root Mean Square Fluctuations (RMSF) of the positions of the atoms during simulations. The ADPs define the $3\times3$ symmetric atomic mean-square displacement tensor $U_{ij}$.
The isotropic displacement parameter can be computed  by $B_{eq}=8\pi^2U_{eq}=\frac{8}{3}\pi^2(U_{11}+U_{22}+U_{33})$.
As $U_{ij}$ are tensors, the comparison of their experimental with simulated values is more complex than with the isotropic ones, so the six independent elements of the symmetric tensor can be compared in different ways, as described by Yang and cols\cite{Yang2009}. Let $U_{ij}$ and $V_{ij}$ be the two tensors to compare, a clear way to do so is to compute the normalized correlation coefficient $ncc(U_{ij},V_{ij})$, defined as:
\begin{equation}
 \label{eq:eq_3}
ncc(U_{ij},V_{ij})=\frac{cc[U, (U_{\text{eq}}/V_{\text{eq}})V]}{cc[U,U^{\text{iso}}]cc[V,V^{\text{iso}}]}
\end{equation}

	where $cc[U_{ij},V_{ij}]=\frac{(\det (U^{-1})\det (V^{-1}))^\frac{1}{4}}{\sqrt[]{(\frac{1}{8})\det(U^{-1}+V^{-1})}}$, $U^{\text{iso}}$ and $V^{\text{iso}}$ are diagonal matrices that describe a pair of isotropic atoms, with $U_{11}^{\text{iso}} =U_{22}^{\text{iso}}=U_{33}^{\text{iso}} = U_{\text{eq}} = \Tr(U_{ij})/3$ and similarly for $V^{\text{iso}}$ and $V_{\text{eq}}$.
    
    The normalized correlation coefficient $ncc$ will have the following values:
\begin{itemize}
    \item $ncc > 1$ if two atoms described by $U$ and $V$ are more similar to each other than to an isotropic atom.
	\item $ncc \leq 1$ otherwise. 
\end{itemize}

With $ncc$, we can compare the size, orientation, and direction of two tensors. If we calculate the ratio of how many atoms in a structure have their $ncc$ values
larger than 1 and the total number of atoms, and express it as a percentage, we can give a good measure of the quality of an anisotropic B{--}factor prediction.

\section{Results and discussion}

\begin{figure}[ht]
\centering
\includegraphics{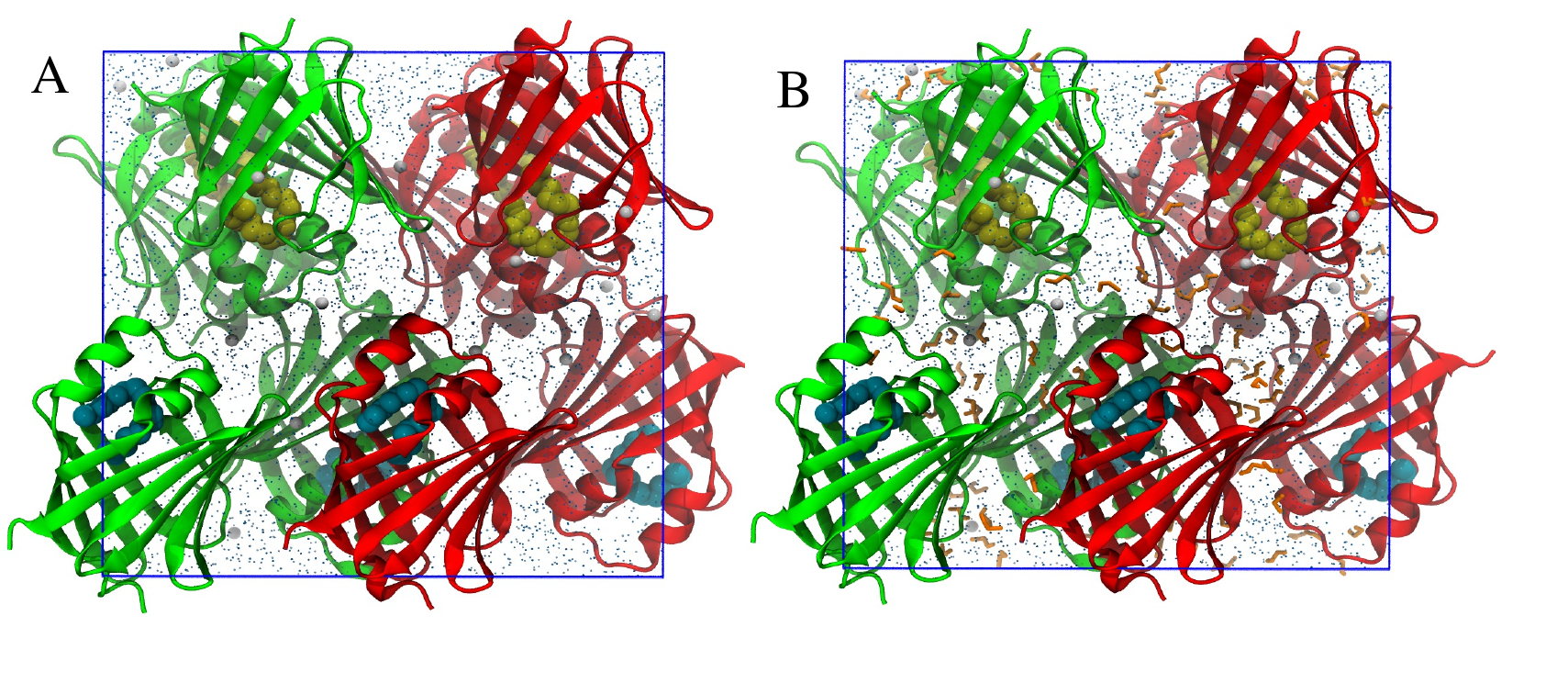}
\caption{Simulated crystal of the heart fatty acid binding proteins, two unit cells containing four proteins each are arranged in a $2\times1\times1$ layout. In green and red the four proteins colored by unit cell. 
The palmitic and oleic acid molecules are represented in cyan and yellow, respectively. Water molecules are indicated using blue spheres. In white spheres Na$^+$ ions are represented. A and B, initial atomic coordinates for the protein crystal in water and in an ethanol(orange):water ratio of 1:37, respectively. }
\label{Fig_1}
\end{figure}

MD simulations were performed using a solvated unit cell model of crystalline H-FABP consisting of two unit cells in a $2\times1\times1$ layout (See Figure \ref{Fig_1}). Analyzed trajectories were obtained during 500 ns of production for the ensembles NVT with restraints, NVT and NpT without restraints, and NVT with ethanol/water keeping restrained only a C$\alpha$ (See Computational methods).

In all our analysis, we applied both a rotational and a translational fit over the C$\alpha$ atoms into all eight protein molecules of each system in order to reduce the overestimation of the positional fluctuations in the residues \cite{stocker2006molecular}.

\begin{figure}[H]
\centering
\includegraphics{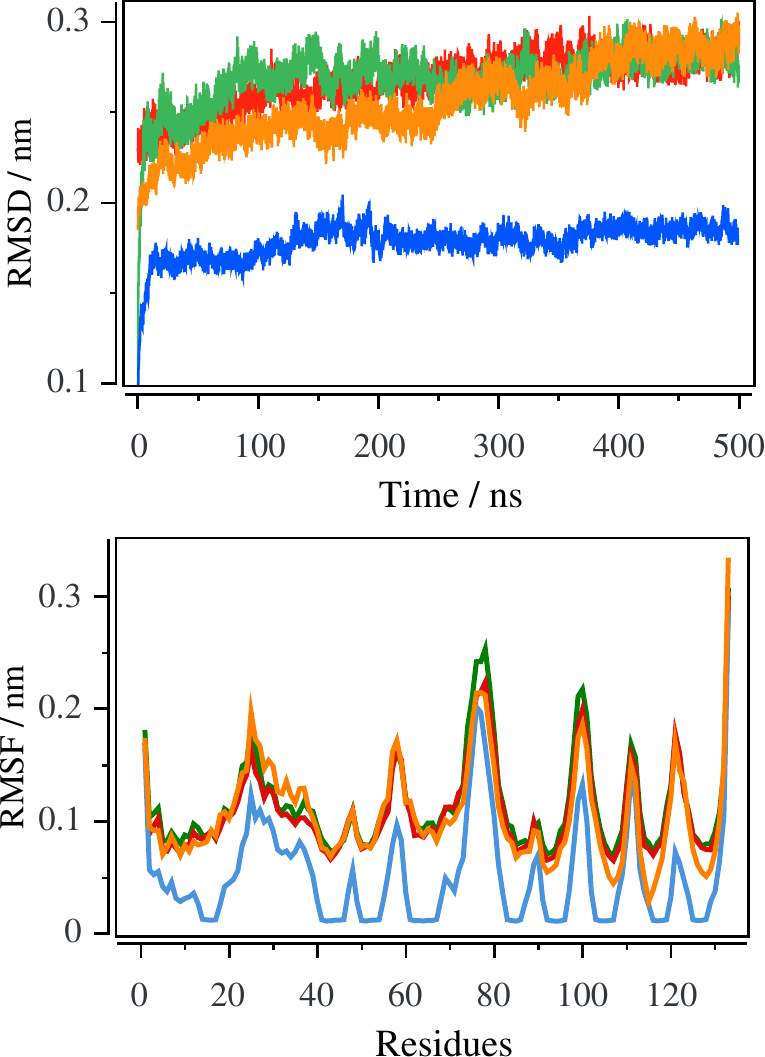}
\caption{Structural and mobility analysis of the H-FABP crystal. 
        In A, Root mean square deviation (RMSD) was calculated aligning each of
        the crystals with the X-Ray and Neutron diffraction structure of H-FABP 
        \cite{howard2016}. In B, root mean square fluctuation (RMSF) obtained by
        averaging the eight of RMSF curves computed on the C$\alpha$ atoms of the 
        individual proteins. In blue, orange, green and red, NVT with position  
        restraints, NVT in ethanol/water, NVT and NpT, respectively.}
\label{Fig_2}
\end{figure}

Initially, we analyze the stability of the system calculating the root mean square deviation (RMSD) of the protein atomic positions and root mean square fluctuation (RMSF) of the positions of the C$\alpha$ atoms in each H-FABP residue. In figure \ref{Fig_2}A, we show that the RMSD in the crystal with ethanol does not converge as fast as in the other systems, becoming stable approximately at the 400 ns (RMSD values around 0.27 nm). 
Predictably, the NVT crystal with positions restrained in forty{--}four of its 
C$\alpha$ atoms (the ones with a B{--}factor lower than a value near 10) shows the lowest RMSD value ($\sim$0.17 nm), while the NVT and NpT systems without position restrain converge quickly with no difference between their RMSD values ($\sim$0.27 nm). 

Likewise, in the RMSF shown in figure \ref{Fig_2}B, we observe that in protein crystals at different conditions the movement throughout the systems tends to have similar dynamics, and despite the restraint in the C$\alpha$ atoms, the crystals show a qualitative correlation in their motions, indicating that the position restrain of the atoms with the lowest B{--}factor is a good strategy to maintain the geometry of the crystal without losing the relevant motions in the proteins. 

\subsection{Essential motions}

To better understand the important protein movements occurred in the simulations, we analyzed the trajectories of the C$\alpha$ atoms from H-FABP crystal using principal component analysis (PCA). Thus, it is possible to detail the direction and amplitude of movements which are relevant for the functioning of the proteins \cite{Amadei1993}.

The C$\alpha$ covariance matrices for the eight H-FABP molecules into the crystal were diagonalized to obtain the eigenvectors and their associated eigenvalues. Subsequently, the trajectory for each system was projected onto the eigenvectors to obtain the principal components. 

\begin{figure}[H]
\centering
\includegraphics{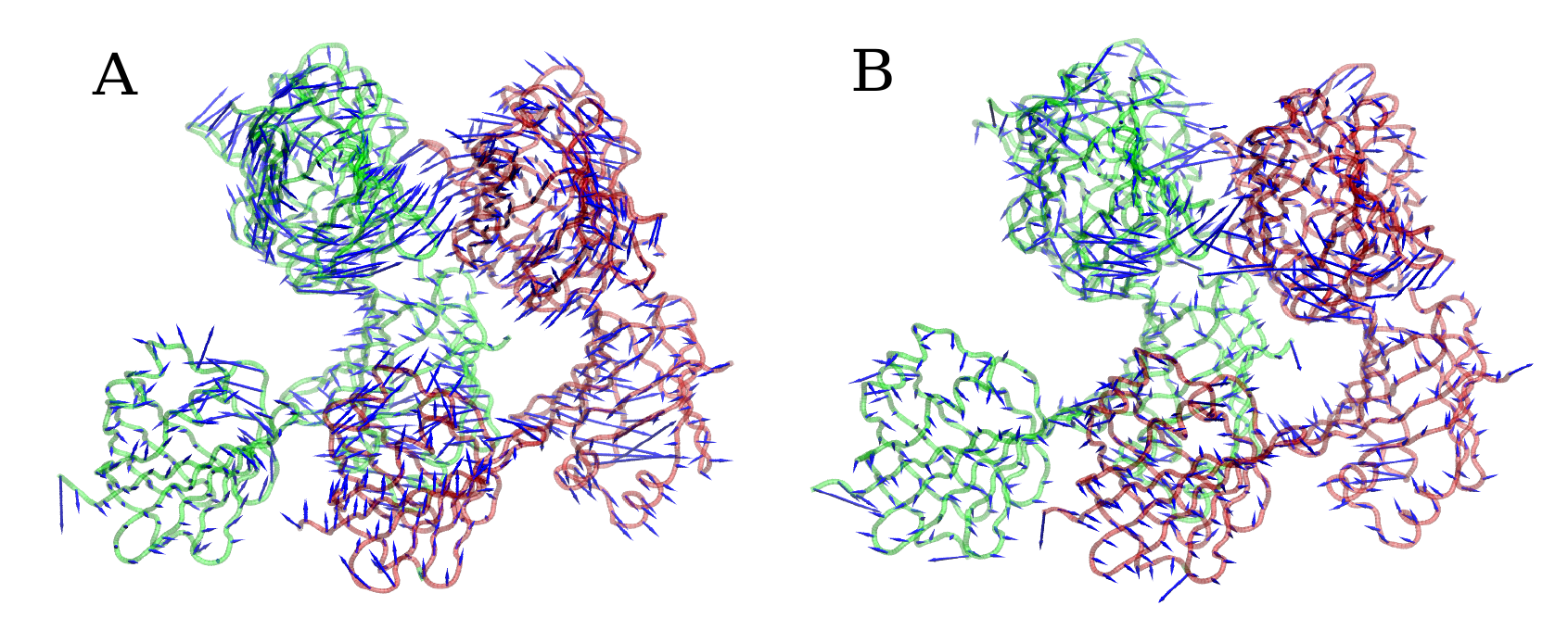}
\caption{View of the essential movements in the H-FABP crystal for the  first principal component. The arrows represent the positional fluctuations, their orientation indicating the direction of motion of the C$\alpha$ atom to which they are attached and their length indicating the amplitude of this motion. In A and B, NVT and NVT with positional restraints, respectively.}
\label{Fig_3}
\end{figure}

In our analysis, we observed that the top ten components with largest amplitudes, represent 55.09\% of the movements for NpT, 51.69\% for NVT, 38.57\% for NVT with position restraints and finally, 59.46\% for NVT with ethanol/water. Interestingly, for NVT ensemble with position restraints, the top components with largest amplitudes represent the lowest percentage of movements in crystal even when compared with the first one hundred components from the other systems (See Figure 2S in Supporting Information). In this particular case, the position restraints minimize the mobility of atoms, as shown in the figure \ref{Fig_3}, at the same time they reduce the fluctuations in the unrestricted atoms in the protein, \emph{i.e.}, the total atomic fluctuation in the crystal is restricted (See Figure 3S in Supporting Information).

\begin{figure}[H]
\centering
\includegraphics{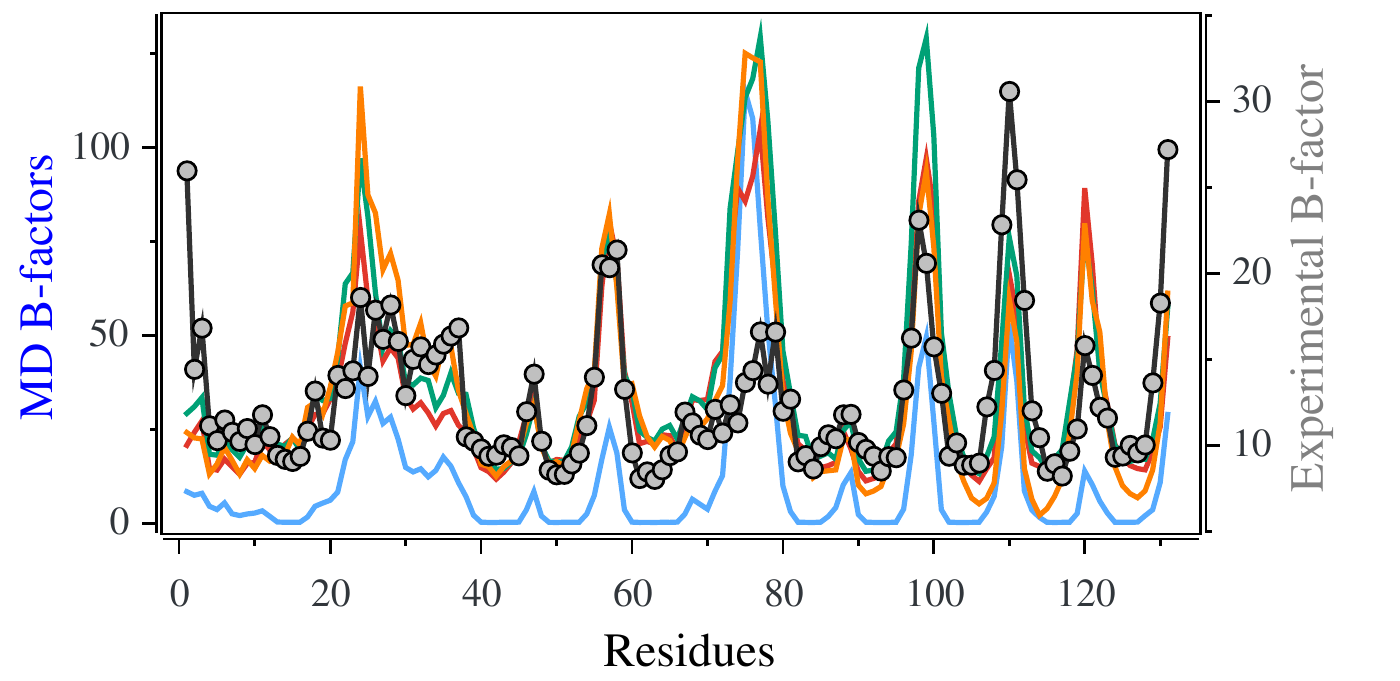}
\caption{Simulation B{--}factors in comparison to experimental B{--}factor of 5CE4 \cite{howard2016}. 
         The experimental B{--}factor is shown as a line with gray dots. The simulation B{--}factors were obtained by averaging the eight B{--}factor curves for the individual proteins in the crystal. NVT with position restraints, NVT in ethanol/water, NVT and NpT, are shown as a line blue, orange, green and red, respectively. The C$\alpha$ atoms were used in this analysis.}
\label{Fig_4}
\end{figure}

Moreover, in the visual inspection of figure \ref{Fig_3}, the H-FABP molecules without restraints show a cooperative movement, which is mitigated when the atomic movement in the crystal is restricted. Thus, to analyze the degree of stability of the crystal in the conformational space during the simulation, the local flexibility of each H-FABP molecule was analyzed by calculating the per-residue B{--}factors (C$\alpha$ B{--}factor), before being averaged over the H-FABPs both unit cells, to be subsequently compared with the crystallographic B{--}factor (See Figure \ref{Fig_4}). Thus, the average C$\alpha$ B{--}factors were calculated as \cite{kuzmanic2010}: 

\begin{equation}
 \label{eq:eq_4}
 B = \frac{8\pi^2}{3} \text{RMSF}^2,
\end{equation}

The simulation B{--}factors analysis showed greater local flexibility. However, although there is an overestimation of the calculated B{--}factors, except for NVT with position restraints, a good qualitative correlation of the simulated with the experimental B{--}factors is evident.

In addition, we can analyze the sampling convergence computing the root mean square inner product (RMSIP) as a measure of similarity between subspaces of each system \cite{papaleo2009}. Thus, the overlap ($O$) between a given PC vector $Y$ and another PC vector $X$ is evaluated by their normalized projection \cite{batista2011},

\begin{equation}
 \label{eq:eq_5}
 O = \frac{Y\cdot X}{\parallel Y \parallel \parallel X \parallel},
\end{equation}

where $Y$ and $X$ are PC vectors from two trajectories at different ensembles.

Our definition of essential subspace of each system was defined by the one hundred eigenvectors with higher eigenvalues, which represented 3.31\% of the total configurational spaces ($3N = 3192$), recovered around 82.04\% (NpT), 81.86\% (NVT), 74.34\% (NVT{--}restraints) and 83.43\% (ethanol/water) of the total motions in the crystal. Thus, the overlap between the essential subspace of two different groups was obtained from the RMSIP as,

 \begin{equation}
  \label{eq:eq_6}
  \text{RMSIP} = \frac{1}{100} \left( \sum_{i=1}^{100} \sum_{j=1}^{100} (n_{i}.v_{j}) \right)^{1/2},
 \end{equation}

where $n_{i}$ and $v_{j}$ are the eigenvectors of the subspaces to be compared. RMSIP ranges from 0 to 1. A perfect match of the sampled subspaces yields an overlap value of 1.

\begin{table}[ht]
\centering
\caption{The root mean square inner products between the one hundred eigenvectors with largest eigenvalues.}
\label{table-1}
\begin{tabular}{lllcc}
\hline
                   & \multicolumn{4}{c}{Root mean square inner product}                                                                               \\ \cline{2-5} 
Eigenvector system & NpT                     & NVT                       & \multicolumn{1}{l}{NVT(restraint)} & \multicolumn{1}{c}{NVT(Ethanol/water)} \\ \hline
NpT                & \multicolumn{1}{c}{1.0} & \multicolumn{1}{c}{0.761} & 0.636                             & 0.736                                  \\
NVT                &                         & \multicolumn{1}{c}{1.0}   & 0.630                             & 0.748                                  \\
NVT(restraint)      &                         &                           & 1.0                               & 0.656                                  \\
NVT(Ethanol/water) &                         &                           & \multicolumn{1}{l}{}              & 1.0                                    \\ \hline
\end{tabular}
\end{table}

According to our analysis, we observed that independently of the ensemble simulated, the RMSIP values were around 0.63{--}0.76, indicating global patterns of correlated movements and a satisfactory overlap between essential subspaces of each system \cite{amadei1999}. Moreover, the similarity of essential subspaces tends to be the lower (between 0.63{--}0.65) when the systems NpT, NVT, and NVT in ethanol/water are overlapped with the NVT-position-restraints system (See Table \ref{table-1}). However, NVT with position restraints represented quantitative sampling that better allowed the study of the atomic fluctuations in the crystal, in agreement with experimental B{--}factor (Figure \ref{Fig_4}). 

Finally, with the aim of analyzing the effect of crystallographic packing on the mobility of the residues, we simulated a single protein solution in 500 ns using an NpT ensemble without restrictions, following the minimization and equilibration protocol described in Computational methods section.

\begin{figure}[h]
\centering
\includegraphics{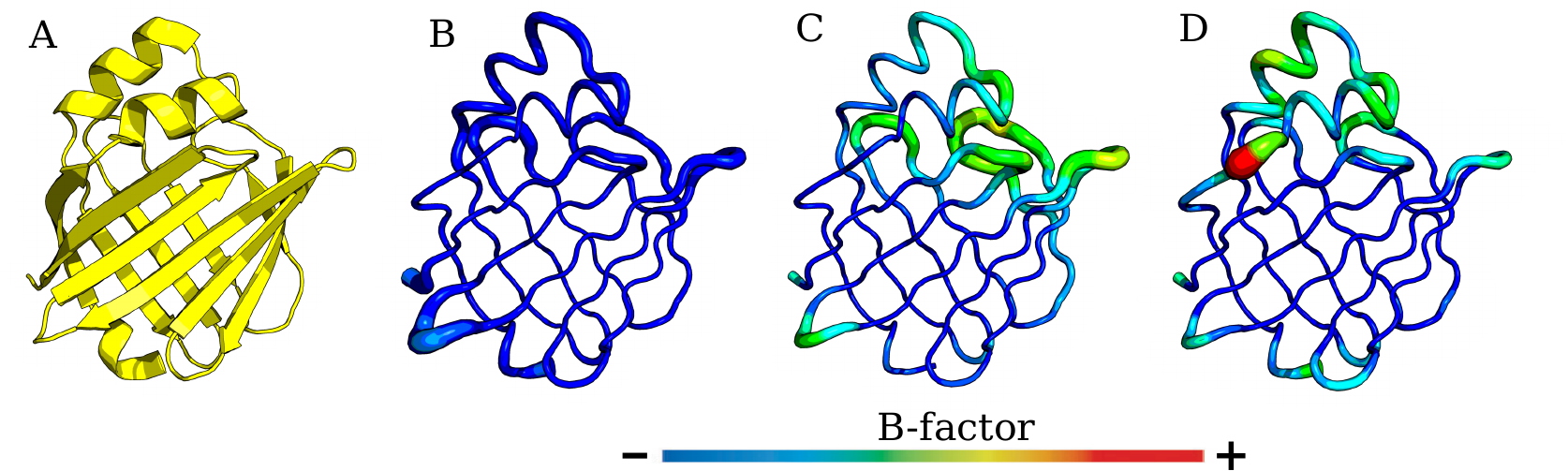}
\caption{Isotropic B{--}factor of H-FABP in NpT ensemble as color code. In A, H-FABP structure from PDB 5CE4. In B{--}D, B{--}factors of experimental H-FABP, simulated protein in crystal and simulated protein in solution, respectively. H-FABP structure in A is shown to understand the configurations represented as B{--}factors.}
\label{Fig_5}
\end{figure}

As seen in Figure \ref{Fig_5}, when the H-FABP is subjected to the crystallographic packing the fluctuation in its movements is reduced in relation to the H-FABP in solution, observing in addition, fluctuations that differ between regions of the proteins (See figure \ref{Fig_5} C and D). Moreover, as observed in figure \ref{Fig_4}, the experimental B{--}factor is smaller in relation to the simulated ones, keeping a greater similarity with the global movements observed in simulated H-FABP in a crystallographic packing.

\begin{figure}[H]
\centering
\includegraphics[scale=0.80]{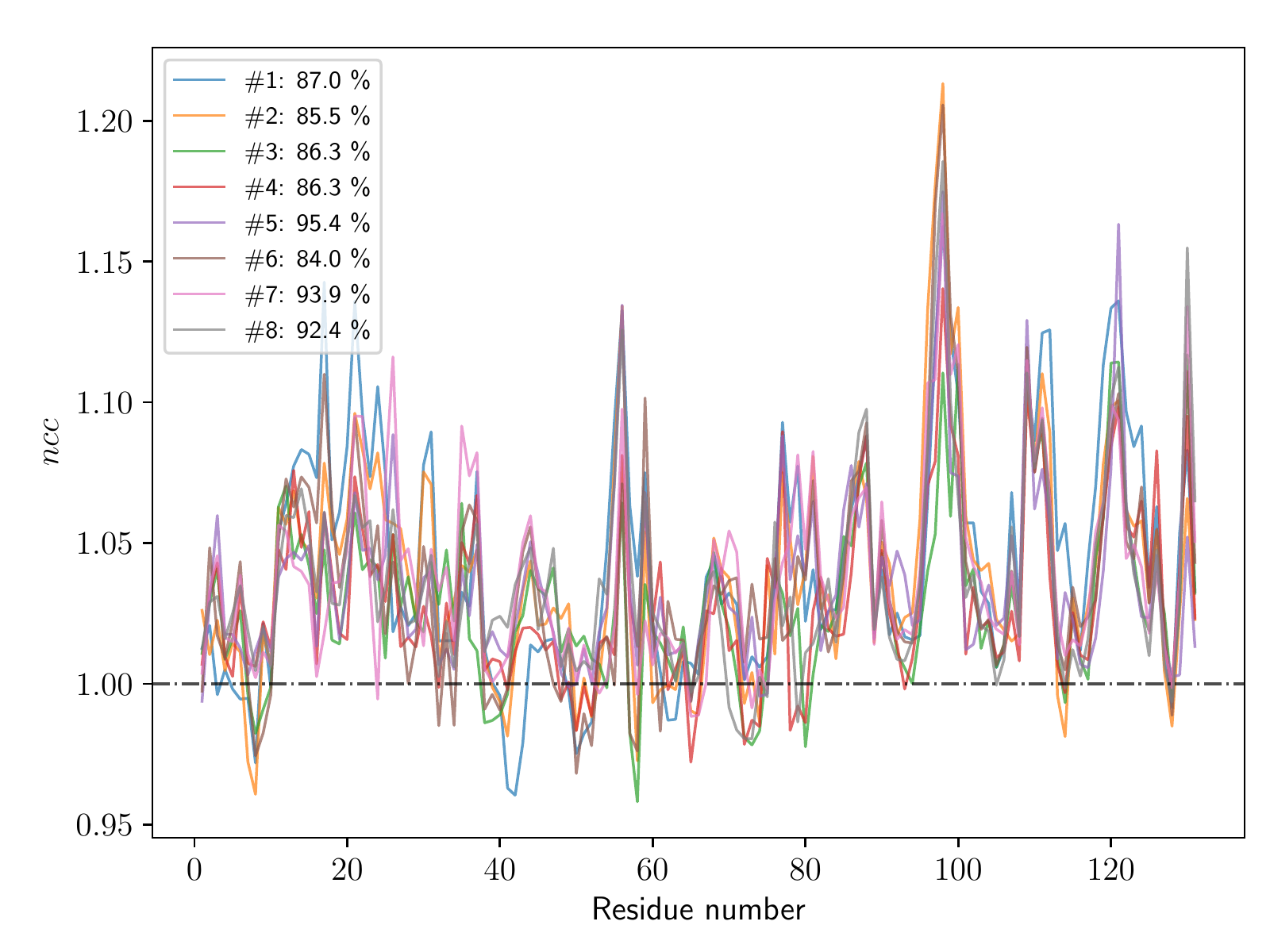}
\caption{$ncc$ values for all eight H-FABP structures from crystal simulation in an ethanol-water solution, compared to the experimental structure as reference. The percentages in the legend indicate the number of residues with $ncc > 1$ relative to the protein length.}
\label{Fig_ncc}
\end{figure}

The normalized correlation coefficients $ncc$ are calculated to compare the experimental anisotropic temperature factor with those predicted by our simulations, in order to get a clear picture of the quality of the MD trajectories obtained, that intend to represent a true crystal system. From Figure \ref{Fig_ncc} we can see that in each simulated chain, the percentage of residues with $ncc > 1$ (which means that the prediction is good) is high, with an averaged value of $88.85$ \%. These results show that there exists high similarity between the experimental anisotropic B-factors and the ones predicted by simulation. 

\section{Conclusion and perspectives}
In the present work, we have explored different strategies to simulate a protein crystal starting from high resolution coordinates obtained at room temperature, which allowed us to build an accurate initial model.
We have done simulations at constant pressure and at constant volume, and we have also modified the number of atoms with restrains to maintain the structure of the crystal. 

These strategies allowed us to evaluate the motions of H-FABP in a confined crystalline environment and in solution, observing how the restriction in the atomic position influences the global motions of the system.

We simulated the crystal system at constant pressure and volume, and despite the good agreement of simulated and experimental B{--}factors (Figure \ref{Fig_4}), the edge proteins showed a high fluctuation in some of its residues (Figure 3S in Supporting Information). However, the unit cells edge volume is well reproduced, indicating that H-FABP packing is described correctly (Figure \ref{Fig_2}A and \ref{Fig_3}). We then proceed to run a simulation at constant volume restricting 44 C$\alpha$ atoms per protein, which decrease the total fluctuation of the H-FABP, but showed a pattern of fluctuations and orientation in their motions consistent with the experimental data (See Figure \ref{Fig_2}B and \ref{Fig_4}). So we reintroduced a restriction (1C$\alpha$/protein) in the ethanol/water system searching for a better match between freedom of movement and simplicity of analysis.

In our analysis, we consider to use the essential dynamics for the calculation of the PCs \cite{daidone2012}. Since the positional fluctuations are confined to a crystallographic cell, the essential dynamics gives a correct description of the motions when its amplitude is small enough (Figure 2S and 3S in Supporting Information). In Table \ref{table-1}, the cross-correlations in the atomic displacements by system indicate collective motion and are, therefore, of potential relevance to H-FABP function \cite{meinhold2005}. 

The results presented here are remarkable considering that a direct comparison between X{--}ray diffraction and MD simulation is difficult, due to the huge differences in the statistical sampling of both techniques. 

A typical experimental X{--}ray data collection is in the order of hundreds of seconds and may involve billions of unit cells. In contrast, in the current computational availability, MD simulations may be extended during microseconds over a small number of unit cells. 
Despite these limitations in the computational modeling, Molecular Dynamic simulations help us to recover part of the information lost in the experiment, introduce movement and therefore the temporary dimension in the atoms positions, reveal the microstates lost in the averaging process, and let us explore the restrictions to the normal movement of the protein due to confinement. All of this enriches the interpretation of the structure from a biological point of view.

The few works carried out so far in this field of MD simulations of crystals are not totally detailed. We hope that this work will help to draw attention to this point, and to clarify it for future studies.

\begin{acknowledgement}

Support of this work by Consejo Nacional de Investigaciones Cient\'{\i}ficas y T\'{e}cnicas (CONICET) and Universidad Nacional de La Plata of Argentina is greatly appreciated. E.I.H. and C.M.C. are members of CONICET - Argentina. H.A.A. is teaching researcher from UNLP and Y.R.E. was supported by the CONICET.

\end{acknowledgement}

\bibliography{achemso-demo}

\begin{suppinfo}
\begin{figure}[ht]
\centering
\includegraphics[scale=0.60]{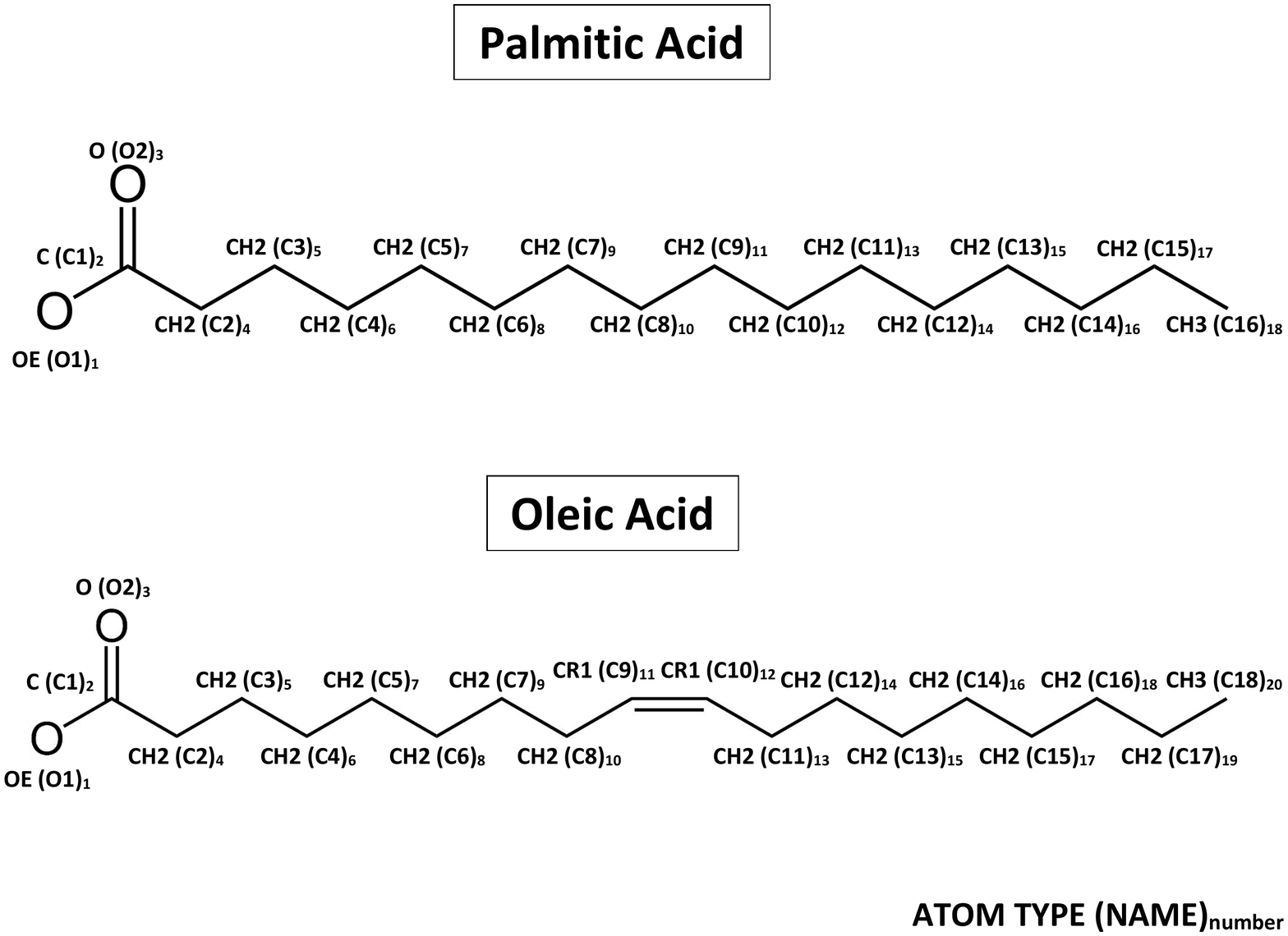}
\caption{Topologies labeling and numbering for palmitic acid and oleic acid that were incorporated in GROMOS96 54A7 force field. Parameter values can be seen in the corresponding topology files also incorporated as Supporting Information (files \text{pal-54a7.itp-topology.txt} and \text{ola-54a7.itp-topology.txt)}.}
\label{Fig_1S}
\end{figure}

\begin{longtable}[c]{c c c c c}
\caption{Temperature factor (B{--}factor) values for the Carbon (C), Nitrogen (N) and C{$\alpha$} atoms in the H{--}FABP (PDB code 5CE4). In bold, the C$\alpha$ atoms from residues with a B{--}factor lower than a value near 10.} \\ \toprule
\textbf{Amino acid} & \textbf{Residue} & \textbf{C} & \textbf{N} & \textbf{C$\alpha$} \\ \toprule 
\endfirsthead
 
\multicolumn{5}{c}{} \\
\caption{Temperature factor (B{--}factor) values for the Carbon (C), Nitrogen (N) and C{$\alpha$} atoms in the H{--}FABP (PDB code 5CE4). In bold, the C$\alpha$ atoms from residues with a B{--}factor lower than a value near 10.}\\ 
\textbf{Amino acid} & \textbf{Residue} & \textbf{C} & \textbf{N} & \textbf{C$\alpha$} \\ \toprule %
\endhead
 
\multicolumn{5}{c}{\textsl{\textit{(Continues on next page)}}}\\ 
\endfoot
 
\multicolumn{5}{c}{\textit{\textsl{(End of the table)}}} \\
\endlastfoot
       
MET & 0  &                       &                       &                        \\
VAL & 1  & 16.31                 & 36.78                 &  25.98                 \\
ASP & 2  & 13.64                 & 14.84                 &  14.43                 \\
ALA & 3  & 14.49                 & 14.72                 &  16.83                 \\
PHE & 4  & 10.09                 & 11.99                 &  11.14                 \\
LEU & 5  & 10.58                 & 10.46                 &  10.28                 \\
GLY & 6  & 9.59                  & 10.47                 &  11.49                 \\
THR & 7  & 10.48                 & 10.26                 &  10.77                 \\
TRP & 8  & 9.81                  & 9.84                  &  10.24                 \\
LYS & 9  & 9.68                  & 9.91                  &  10.99                 \\
LEU & 10 & 10.55                 & 9.96                  &  10.05                 \\
VAL & 11 & 11.36                 & 10.22                 &  11.79                 \\
ASP & 12 & 9.77                  & 10.77                 &  10.51                 \\
SER & 13 & 9.5                   & 9.8                   &  \textbf{9.42}                  \\
LYS & 14 & 8.36                  & 8.68                  &  \textbf{9.19}                  \\
ASN & 15 & 9.08                  & 8.47                  &  \textbf{9.07}                  \\
PHE & 16 & 9.32                  & 8.82                  &  \textbf{9.36}                  \\
ASP & 17 & 10.68                 & 9.97                  &  10.83                 \\
ASP & 18 & 11.33                 & 12.18                 &  13.17                 \\
TYR & 19 & 10.04                 & 10.75                 &  10.43                 \\
MET & 20 & 11.34                 & 9.85                  &  10.31                 \\
LYS & 21 & 13.87                 & 12.19                 &  14.08                 \\
SER & 22 & 13.84                 & 13.2                  &  13.32                 \\
LEU & 23 & 16.63                 & 14                    &  14.35                 \\
GLY & 24 & 17.27                 & 16.75                 &  18.61                 \\
VAL & 25 & 14.87                 & 15.1                  &  14.02                 \\
GLY & 26 & 15.5                  & 16.44                 &  17.87                 \\
PHE & 27 & 14.67                 & 15.93                 &  16.17                 \\
ALA & 28 & 16.68                 & 16.51                 &  18.18                 \\
THR & 29 & 13.91                 & 16.33                 &  16.05                 \\
ARG & 30 & 12.87                 & 13.61                 &  12.9                  \\
GLN & 31 & 14.17                 & 13.32                 &  15.01                 \\
VAL & 32 & 14.57                 & 14.61                 &  15.74                 \\
ALA & 33 & 14.67                 & 13.8                  &  14.71                 \\
SER & 34 & 15.23                 & 13.91                 &  15.27                 \\
MET & 35 & 16.77                 & 14.36                 &  15.9                  \\
THR & 36 & 15.09                 & 15.26                 &  16.37                 \\
LYS & 37 & 14.56                 & 14.99                 &  16.84                 \\
PRO & 38 & 9.53                  & 11.68                 &  10.5                  \\
THR & 39 & 9.86                  & 9.61                  &  10.25                 \\
THR & 40 & 8.78                  & 9.45                  &  \textbf{9.8}                   \\
ILE & 41 & 8.81                  & 9.08                  &  \textbf{9.38}                  \\
ILE & 42 & 8.77                  & 9.15                  &  \textbf{9.44}                  \\
GLU & 43 & 8.8                   & 9.13                  &  \textbf{10.04}                 \\
LYS & 44 & 8.62                  & 9.1                   &  \textbf{9.88}                  \\
ASN & 45 & 8.9                   & 8.8                   &  \textbf{9.4}                   \\
GLY & 46 & 12.82                 & 10.81                 &  11.97                 \\
ASP & 47 & 12.09                 & 12.38                 &  14.15                 \\
ILE & 48 & 9.15                  & 10.66                 &  10.25                 \\
LEU & 49 & 8.39                  & 8.68                  &  \textbf{8.56}                  \\
THR & 50 & 8.59                  & 8.47                  &  \textbf{8.3}                   \\
LEU & 51 & 7.87                  & 8.06                  &  \textbf{8.32}                  \\
LYS & 52 & 8.53                  & 8.52                  &  \textbf{8.93}                  \\
THR & 53 & 9.79                  & 8.72                  &  \textbf{9.56}                  \\
HIS & 54 & 11.49                 & 10.52                 &  11.16                 \\
SER & 55 & 15.69                 & 12                    &  13.97                 \\
THR & 56 & 20.99                 & 17.96                 &  20.51                 \\
PHE & 57 & 19.56                 & 19.93                 &  20.35                 \\
LYS & 58 & 15.82                 & 19.22                 &  21.38                 \\
ASN & 59 & 11.25                 & 14.58                 &  13.27                 \\
THR & 60 & 8.18                  & 9.67                  &  \textbf{9.57}                  \\
GLU & 61 & 7.95                  & 8.32                  &  \textbf{8.08}                  \\
ILE & 62 & 8.04                  & 8.18                  &  \textbf{8.47}                  \\
SER & 63 & 8.72                  & 8.35                  &  \textbf{8.03}                  \\
PHE & 64 & 8.12                  & 8.13                  &  \textbf{8.59}                  \\
LYS & 65 & 8.8                   & 8.87                  &  \textbf{9.41}                  \\
LEU & 66 & 9.75                  & 9.52                  &  \textbf{9.63}                  \\
GLY & 67 & 12.09                 & 10.21                 &  11.94                 \\
VAL & 68 & 10.03                 & 11.23                 &  11.32                 \\
GLU & 69 & 9.54                  & 10.41                 &  10.6                  \\
PHE & 70 & 10.2                  & 9.36                  &  10.35                 \\
ASP & 71 & 11.7                  & 10.88                 &  12.1                  \\
GLU & 72 & 10.71                 & 11.39                 &  10.72                 \\
THR & 73 & 11.82                 & 11.55                 &  12.37                 \\
THR & 74 & 11.73                 & 11.18                 &  11.32                 \\
ALA & 75 & 14.33                 & 12.1                  &  13.67                 \\
ASP & 76 & 15.63                 & 13.96                 &  14.36                 \\
ASP & 77 & 16.67                 & 15.09                 &  16.61                 \\
ARG & 78 & 13.55                 & 13.78                 &  13.57                 \\
LYS & 79 & 13.82                 & 13.85                 &  16.6                  \\
VAL & 80 & 11.54                 & 12.61                 &  11.99                 \\
LYS & 81 & 10.24                 & 11.66                 &  12.7                  \\
SER & 82 & 8.75                  & 9.75                  &  \textbf{9.05}                  \\
ILE & 83 & 8.66                  & 8.92                  &  \textbf{9.41}                 \\
VAL & 84 & 8.61                  & 8.86                  &  \textbf{8.64}                  \\
THR & 85 & 9.19                  & 9.39                  &  \textbf{9.93}                  \\
LEU & 86 & 10.17                 & 10.77                 &  10.62                 \\
ASP & 87 & 9.28                  & 9.92                  &  10.39                 \\
GLY & 88 & 11.61                 & 10.12                 &  11.79                 \\
GLY & 89 & 11.59                 & 10.84                 &  11.81                 \\
LYS & 90 & 9.56                  & 10.53                 &  10.18                 \\
LEU & 91 & 9.07                  & 9.24                  &  \textbf{9.77}                  \\
VAL & 92 & 8.95                  & 9.18                  &  \textbf{9.37}                  \\
HIS & 93 & 8.93                  & 8.37                  &  \textbf{8.51}                  \\
LEU & 94 & 9.04                  & 9.11                  &  \textbf{9.35}                  \\
GLN & 95 & 10.13                 & 9.17                  &  \textbf{9.29}                  \\
LYS & 96 & 13.26                 & 11.25                 &  13.23                 \\
TRP & 97 & 18.28                 & 14.3                  &  16.25                 \\
ASP & 98 & 21.03                 & 21.25                 &  23.11                 \\
GLY & 99 & 19.08                 & 20.59                 &  20.6                  \\
GLN & 100 &13.14                 & 16.35                 &  15.74                 \\
GLU & 101 &11.19                 & 12.69                 &  13.04                 \\
THR & 102 &9.36                  & 10.1                  &  \textbf{9.38}                  \\
THR & 103 &9.35                  & 9.73                  &  \textbf{10.15}                 \\
LEU & 104 &8.64                  & 8.9                   &  \textbf{8.85}                  \\
VAL & 105 &8.29                  & 8.28                  &  \textbf{8.85}                  \\
ARG & 106 &8.53                  & 8.27                  &  \textbf{8.98}                  \\
GLU & 107 &11.67                 & 10.42                 &  12.25                 \\
LEU & 108 &16.29                 & 12.41                 &  14.37                 \\
ILE & 109 &23.05                 & 19.12                 &  22.84                 \\
ASP & 110 &27.54                 & 25.91                 &  30.6                  \\
GLY & 111 &22.27                 & 26.52                 &  25.46                 \\
LYS & 112 &14.79                 & 19.52                 &  18.44                 \\
LEU & 113 &11.22                 & 12.38                 &  12.02                 \\
ILE & 114 &9.3                   & 10.98                 &  10.46                 \\
LEU & 115 &8.16                  & 9.16                  &  \textbf{8.5}                   \\
THR & 116 &8.11                  & 8.45                  &  \textbf{8.94}                  \\
LEU & 117 &8.7                   & 8.11                  &  \textbf{8.23}                  \\
THR & 118 &9.77                  & 9.16                  &  \textbf{9.67}                  \\
HIS & 119 &11.9                  & 10.5                  &  10.96                 \\
GLY & 120 &15.25                 & 14.43                 &  15.81                 \\
THR & 121 &13.36                 & 14.67                 &  14.08                 \\
ALA & 122 &11.53                 & 12.35                 &  12.23                 \\
VAL & 123 &11.35                 & 11.53                 &  11.59                 \\
CYS & 124 &8.66                  & 9.99                  &  \textbf{9.31}                  \\
THR & 125 &9.57                  & 8.92                  &  \textbf{9.39}                  \\
ARG & 126 &10.09                 & 9.31                  &  \textbf{10.01}                 \\
THR & 127 &9.27                  & 9.6                   &  \textbf{9.57}                  \\
TYR & 128 &11.09                 & 9.66                  &  10.03                 \\
GLU & 129 &12.79                 & 11.88                 &  13.64                 \\
LYS & 130 &19.1                  & 15.02                 &  18.28                 \\
GLU & 131 &26.81                 & 20.68                 &  27.22                      \\
ALA & 132 &                      &                       &                       
\end{longtable}

\begin{figure}[ht]
\centering
\includegraphics{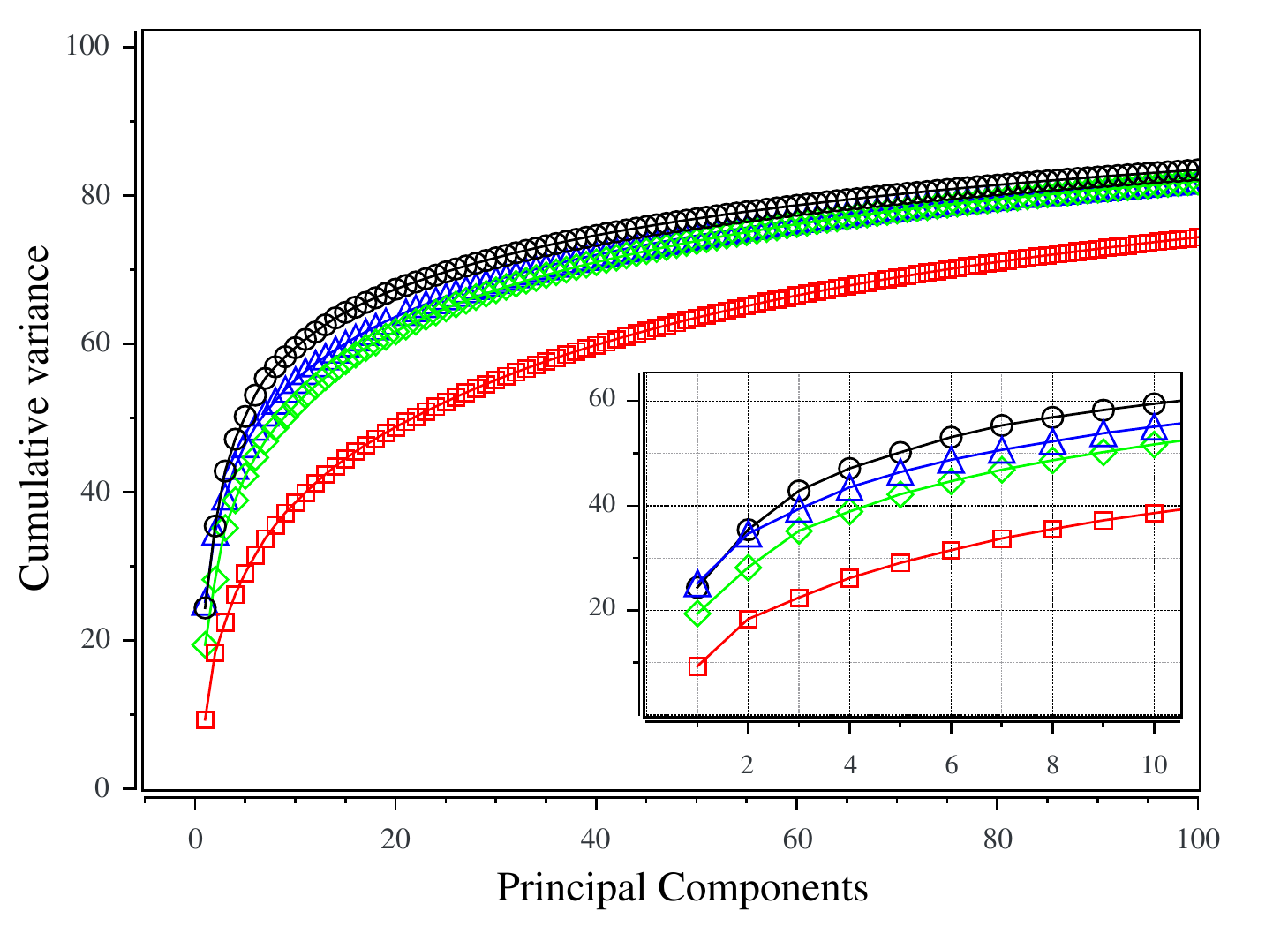}
\caption{Cumulative percentages of total fluctuations (variance) capture by the first 100 PCs. The inset shows a zoomed in the first 10 PCs.}
\label{Fig_2S}
\end{figure}

\begin{figure}[ht]
\centering
\includegraphics{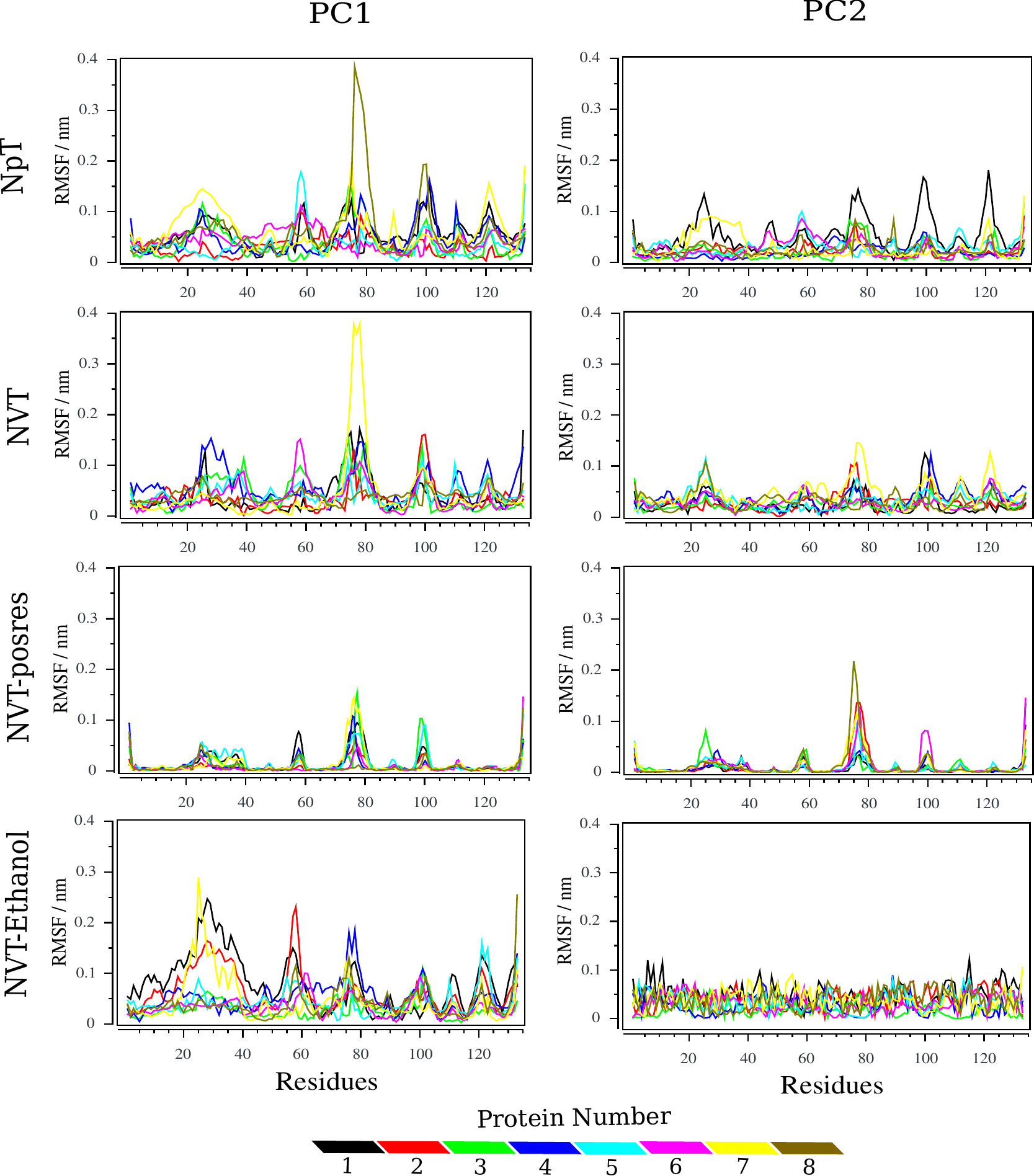}
\caption{RMSF of each H{--}FABP on the crystal. The trajectories of each conditions was projected onto the two first principal components.}
\label{Fig_3S}
\end{figure}

\end{suppinfo}

\end{document}